\documentclass[aps,10pt,pra,twocolumn,showpacs,preprintnumbers,amsmath,amssymb,nofootinbib,floatfix]{revtex4-1}

\usepackage{amsmath}
\usepackage{graphicx}
\usepackage{multirow} 
\usepackage{array} 
\usepackage{longtable} 
\usepackage{bigstrut} 
\usepackage{verbatim}

\newcommand{\gom}{g_{\mathrm{om}}}
\newcommand{\gomC}{g_{\mathrm{C}}}
\newcommand{\inR}{\mathrm{in,}R}
\newcommand{\outR}{\mathrm{out,}R}
\newcommand{\inL}{\mathrm{in,}L}
\newcommand{\outL}{\mathrm{out,}L}
\newcommand{\inV}{\mathrm{in,}V}
\newcommand{\outV}{\mathrm{out,}V}
\newcommand{\inin}{\mathrm{in}}
\newcommand{\out}{\mathrm{out}}
\newcommand{\w}{\omega}
\newcommand{\wm}{\omega_m}
\newcommand{\we}{\omega_{\mathrm{eff}}}
\newcommand{\Gm}{\Gamma_m}
\renewcommand{\k}{\kappa}
\newcommand{\D}{\Delta}
\newcommand{\Gopt}{\Gamma_\mathrm{opt}}
\newcommand{\opt}{\mathrm{opt}}

\newcommand{\detec}{\mathrm{det}}

\newcommand{\ext}{\mathrm{ext}}
\newcommand{\tot}{\mathrm{tot}}
\newcommand{\ho}{\mathrm{ho}}

\newcommand{\therm}{\mathrm{th}}
\newcommand{\ideal}{\mathrm{ideal}}
\newcommand{\hatt}[1]{\hat{\tilde{#1}}}
\newcommand{\Haalph}{\mathbf{H}_{\alpha}}
\newcommand{\Haeta}{\mathbf{H}_{\eta}}
\newcommand{\Hpalph}{H_{+\alpha}}
\newcommand{\Hmalph}{H_{-\alpha}}
\newcommand{\Hpeta}{H_{+\eta}}
\newcommand{\Hmeta}{H_{-\eta}}
\newcommand{\HaSS}{\mathbf{H}_\alpha^{\mathrm{SS}}}
\newcommand{\HaMT}{\mathbf{H}_\alpha^{\mathrm{MT}}}
\newcommand{\Fa}{\mathbf{F}_a}
\newcommand{\Fb}{\mathbf{F}_b}
\newcommand{\Dopt}{D_\mathrm{opt}}
\newcommand{\Copt}{C_\mathrm{opt}}

\newcommand{\ftmatrix}[1]{
	\begin{pmatrix}\hatt{#1}_+ \\ 
			\hatt{#1}_- \end{pmatrix}
}
\newcommand{\ftmatrixIn}[1]{
	\begin{pmatrix}\hatt{#1}_{\inin+} \\ 
			\hatt{#1}_{\inin-} \end{pmatrix}
}

\newcommand{\ftmatrixAP}[1]{
	\begin{pmatrix}\hatt{#1}_{\mathrm{in},A} \\ 
			\hatt{#1}_{\mathrm{in},P} \end{pmatrix}
}
\newcommand{\ftmatrixAPin}[1]{
	\begin{pmatrix}\hatt{#1}_{\mathrm{in},jA} \\ 
			\hatt{#1}_{\mathrm{in},jP} \end{pmatrix}
}
\newcommand{\ftmatrixAPout}[1]{
	\begin{pmatrix}\hatt{#1}_{\mathrm{out},jA} \\ 
			\hatt{#1}_{\mathrm{out},jP} \end{pmatrix}
}

\begin{document}

\title{Linear Amplifier Model for Optomechanical Systems 
}


\author{Thierry Botter$^{1}$}
\email{tbotter@berkeley.edu}
\author{Daniel W. C. Brooks$^{1}$}
\author{Nathan Brahms$^{1}$}
\author{Sydney Schreppler$^{1}$}
\author{Dan M. Stamper-Kurn$^{1,2}$}
\email{dmsk@berkeley.edu}
\affiliation{
$^1$Department of Physics, University of California, Berkeley, CA, 94720, USA \\
$^2$Materials Sciences Division, Lawrence Berkeley National Laboratory, Berkeley, CA, 94720, USA}

\date{\today}

\begin{abstract}
We model optomechanical systems as linear optical amplifiers. This provides a unified treatment of diverse optomechanical phenomena. We emphasize, in particular, the relationship between ponderomotive squeezing and optomechanically induced transparency, two foci of current research. We characterize the amplifier response to quantum and applied classical fluctuations, both optical and mechanical. Further, we apply these results to establish quantum limits on external force sensing both on and off cavity resonance. We find that the maximum sensitivity attained on resonance constitutes an absolute upper limit, not surpassed when detuning off cavity resonance. The theory is extended to a two-sided cavity with losses and limited detection efficiency. 
\end{abstract}
\maketitle


Cavity optomechanics \cite{Kippenberg2007,Kippenberg2008} describes the macroscale effects of radiation pressure on moveable reflective and refractive media \cite{Braginskii1967, Braginskii1975} with cavity-based optical feedback. Research in the field today is conducted on several fronts, from nano- \cite{Anetsberger2010, Eichenfield2009, Lin2010} and micro-fabricated \cite{Sankey2010, Hofer2010, Wilson2009} devices to 
atomic gases \cite{Murch2008, Brennecke2008b} to kilogram-size mirrors \cite{Corbitt2007, Corbitt2006}, and from microwave to optical frequencies. Efforts to elucidate the classical and quantum nature of optomechanical systems have led to demonstrations of sideband cooling \cite{Elste2009, Marquardt2007, Arcizet2006, Schliesser2006, Gigan2006}, amplification \cite{Ludwig2008, Arcizet2006} and backaction evasion \cite{Hertzberg2010}, observations of the optical spring effect \cite{Sheard2004}, quantum-sensitive force detection \cite{Teufel2009, Anetsberger2009}, explorations of optical nonlinearity and bistability \cite{Purdy2010}, and studies of ponderomotive squeezing \cite{Brooks2011} and classical analogs thereof \cite{Marino2010, Verlot2010}. In the past year, further experimental advances also generated the first ground-state oscillators \cite{Teufel2011, OConnell2010} and led to the observation of mechanically induced optical transparency \cite{Safavi2011, Weis2010}.

To date, these varied research avenues have been modeled individually. Works highlighting a particular aspect of optomechanics are often prefaced by extensive derivations to set the context. This effectively isolates different aspects of the same optomechanical interaction, making it difficult to establish the connections between them.

Here, we present a framework that treats these disparate phenomena in a unified manner. Cavity-mediated interactions between a harmonic oscillator and a circulating light field are modeled as a feedback circuit. This allows the use of concepts from control theory. Optomechanical systems are therefore represented as linear optical and mechanical amplifiers with frequency-dependent gain. We study the amplifier response to optical and mechanical inputs for the general case of a two-sided cavity with losses. Results include a connection between ponderomotive squeezing \cite{Brooks2011, Marino2010, Verlot2010, Fabre1994, Borkje2010, Heidmann1997} and optomechanically induced transparency (OMIT) \cite{Safavi2011, Weis2010, Teufel2010}. The amplifier model is also used to set quantum limits on the transduction of external mechanical drives.

\begin{figure*}
\centering
	\includegraphics [width= 8.9cm] {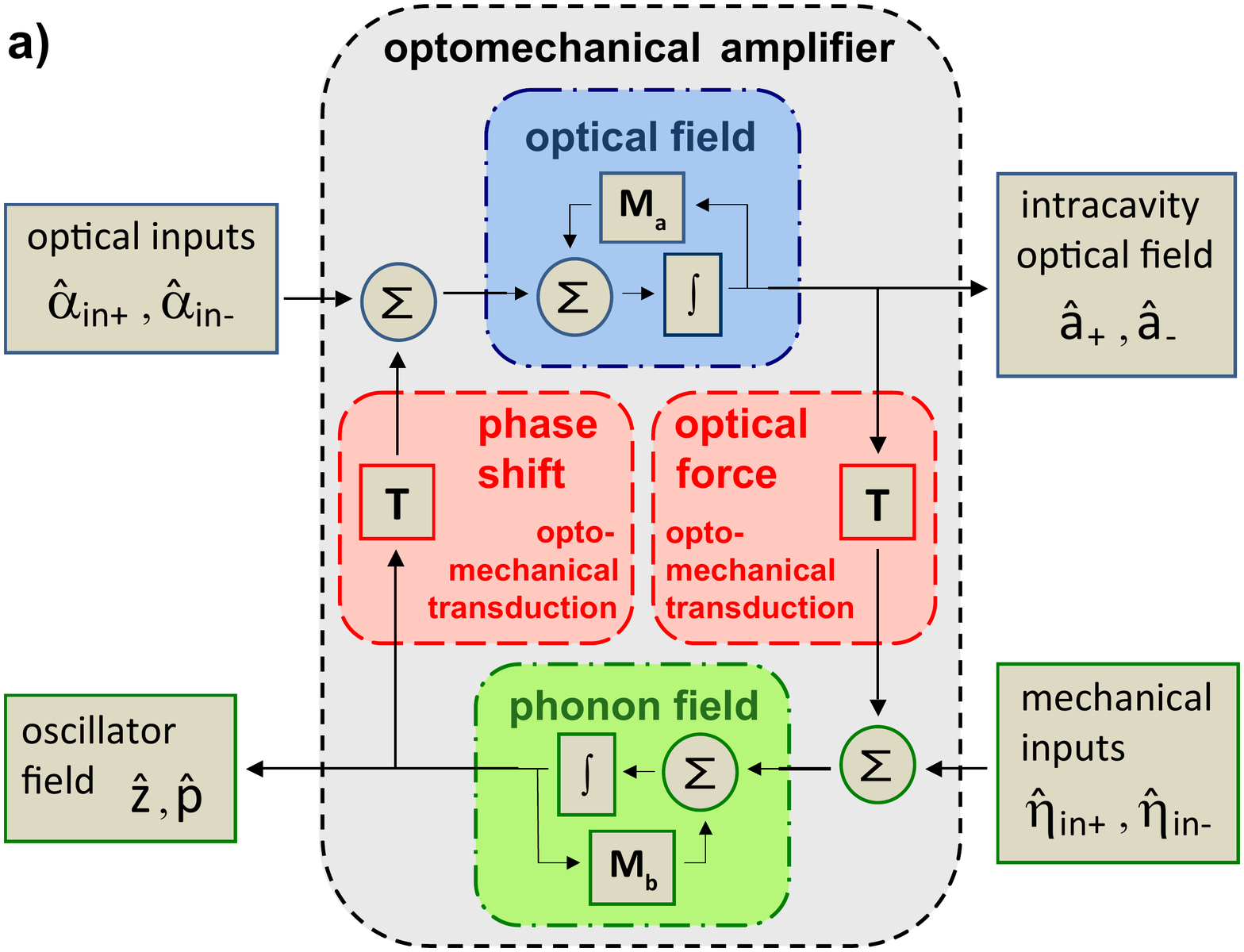} 
	\includegraphics [width= 8.9cm] {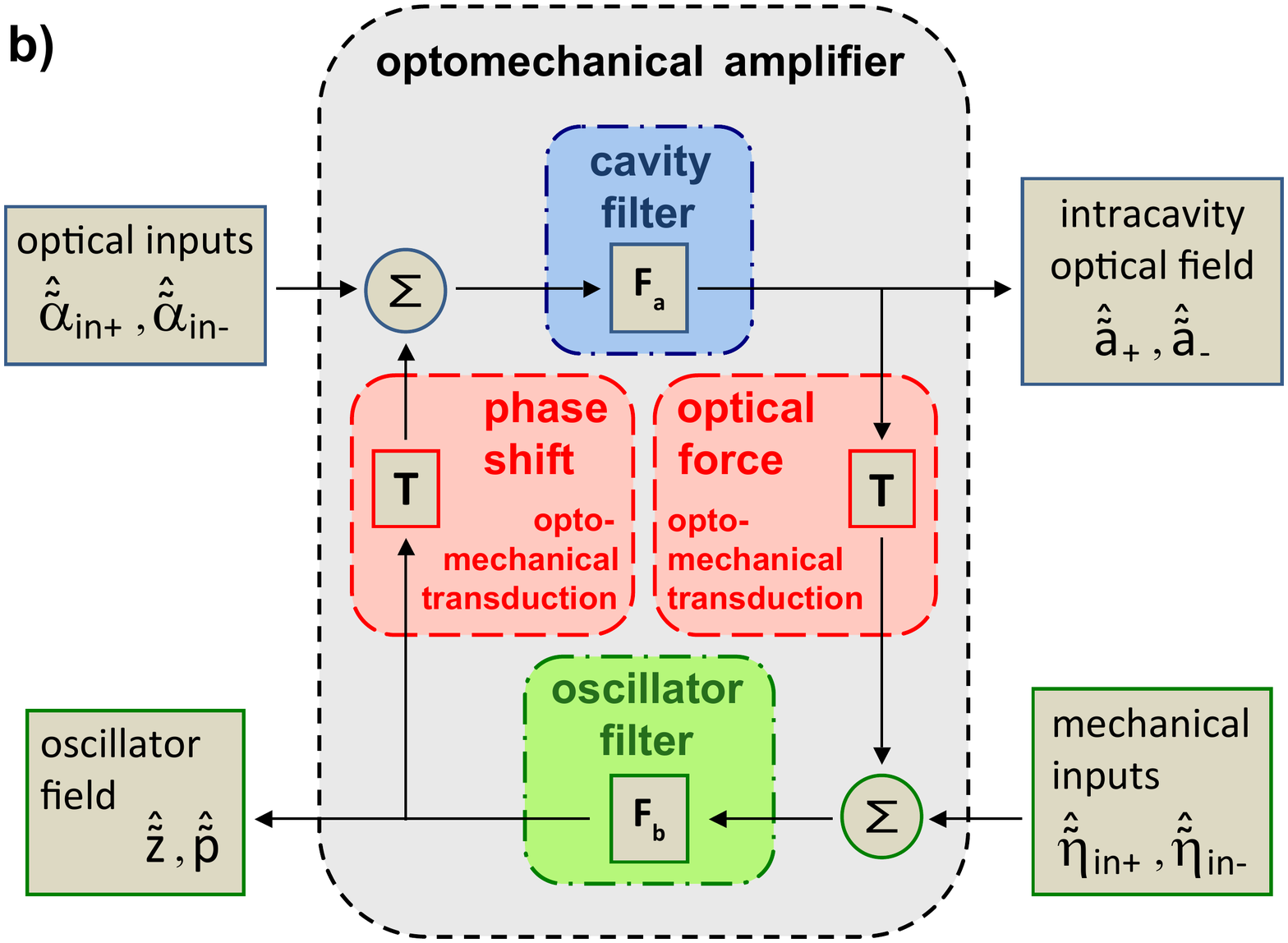} 
	\caption{(Color online) Block diagram model of linear optomechanics in the ($\mathbf{a}$) time domain and ($\mathbf{b}$) frequency domain, as established in Eq.~(\ref{EomTime}) and Eq.~(\ref{EomFreq1}), respectively. The block matrices of Eq.~(\ref{EomTime}) and Eq.~(\ref{EomFreq1}) are shown in their respective circuit. Symbols $\sum$ and $\int$ represent the sum of two inputs and the time integral of an input, respectively. Linear optomechanical feedback takes place via the two center blocks (red), which connect the optical (blue) and mechanical (green) field evolution.} 
	\label{fig:BlockDiagram}
\end{figure*}

\section{Model of Optomechanical Interaction}\label{sect:Model}
We consider a two-sided optical cavity containing one optical element that is moveable and harmonically bound. The element may be one of the cavity mirrors, or an intracavity dispersive element placed at a linear gradient of the light field \cite{Thompson2007}. Light circulating inside the cavity couples parametrically with the position of the harmonic oscillator. The system is described by the Hamiltonian
\begin{equation}\label{H}
	H = \hbar\omega_{c}\hat{a}^\dag\hat{a}+\hbar\omega_{m}\hat{b}^\dag\hat{b}+\hbar \, \gom \, \hat{z}\hat{a}^\dag\hat{a}+\hat{H}_{\kappa}+\hat{H}_{\gamma} \,\,,
\end{equation}
where $\omega_{c}$ ($\omega_{m}$) and $\hat{a}$ ($\hat{b}$) are the resonant frequency and the annihilation operator of the cavity (oscillator) field, respectively. The first two terms of Eq.~(\ref{H}) represent the energy stored in the photon and phonon fields. The final two terms, $\hat{H}_{\kappa}$ and $\hat{H}_{\gamma}$, contain the connections to external photon and phonon baths, respectively. Constant energy offsets are ignored.


The third term in Eq.~(\ref{H}) captures the optomechanical interaction, where $\hat{z}=\hat{b}+\hat{b}^\dag$ is the dimensionless position operator of the oscillator and $\gom$ sets the coupling strength. The interaction has a clear signature of three-wave mixing, with intermodulation between two optical field operators and one mechanical field operator. This fundamentally nonlinear (three-wave) coupling can be linearized about an equilibrium position displacement $z_{0}$ caused by a dominant optical pump field $\hat{a}_{0}e^{-i\omega_{p}t}$ rotating at frequency $\omega_{p}$: 
\begin{eqnarray}\label{operators}
\hat{z}&\to&z_{0}+\hat{z}\,\,, \\
\nonumber
\hat{a}&\to&\left(a_{0}+\hat{a}\right)e^{-i\omega_{p}t}\,\,.
\end{eqnarray}
For simplicity, we set $a_{0}=\sqrt{n}$ to be real, with $n$ being the mean intracavity pump photon number. The resulting linearized optomechanical interaction energy can be written in two parts:
\begin{equation}\label{Hstatic}
H_{\mathrm{static}}=\hbar\gom\left(z_{0}n+n\hat{z}+\sqrt{n}\,
z_{0}(\hat{a}+\hat{a}^\dag)\right)\,\,, 
\end{equation}
describing static changes in the cavity resonance frequency and the oscillator position, and 
\begin{equation}\label{Hdyn}
H_{\mathrm{dyn}}=\hbar\gom\left(\sqrt{n}\,
\hat{z}(\hat{a}+\hat{a}^\dag)\right)\,\,,
\end{equation}
corresponding to linearized dynamics. Eq.~(\ref{Hdyn}) shows that the effective linear coupling between $\hat{z}$ and $\hat{a}$ is mediated by the pump field $\sqrt{n}$, 
with optical sidebands interpreted as signals and phonon modes as idlers. 

The exchange of information between the circulating light field and the oscillator, captured by $H_{\mathrm{dyn}}$, leads to dynamical backaction \cite{Kippenberg2008} or, in the language of control-systems engineering, feedback. To model this feedback, we introduce conjugate quadratures for both field fluctuation operators:
\begin{align}\label{Quadratures}
\hat{b}_{+}&=\hat{z}=\hat{b}+\hat{b}^{\dagger}\,\,, & \hat{b}_{-}&=\hat{p}=i(\hat{b}-\hat{b}^{\dagger})\,\,, \\
\nonumber
\hat{a}_{+}&=\hat{a}+\hat{a}^{\dagger}\,\,, & \hat{a}_{-}&=i(\hat{a}-\hat{a}^{\dagger})\,\,.
\end{align}
Operators $\hat{z}$ and $\hat{p}$ represent the dimensionless position and momentum fluctuation operators of the oscillator, while $\hat{a}_{+}$ and $\hat{a}_{-}$ are the amplitude modulation (AM) and phase modulation (PM) quadratures of the intracavity optical field. Time evolution of these operators is provided by the Heisenberg equations in a frame co-rotating with the pump field:
\begin{eqnarray}\label{EomTime}
\left( \begin{array}{c} 
\dot{\hat{a}}_{+} \\ \dot{\hat{a}}_{-} \\ \dot{\hat{z}} \\ \dot{\hat{p}} 
\end{array} \right)&=&\mathbf{M}
\left( \begin{array}{c} 
\hat{a}_{+} \\ \hat{a}_{-} \\ \hat{z} \\ \hat{p} 
\end{array} \right) + 
\left( \begin{array}{c}
\sqrt{\gamma_{T}}\,\hat{\alpha}_{\inin+} \\
\sqrt{\gamma_{T}}\,\hat{\alpha}_{\inin-} \\
\sqrt{\Gamma_{m}}\,\hat{\eta}_{\inin+} \\ 
\sqrt{\Gamma_{m}}\,\hat{\eta}_{\inin-} 
\end{array} \right)\,\,, 
\end{eqnarray}
where
\begin{eqnarray}\label{M}
\mathbf{M}&=&
\begin{pmatrix} \mathbf M_a & \mathbf T \\ \mathbf T & \mathbf M_b\end{pmatrix} =
\left( \begin{array}{cc|cc} 
-\kappa & \Delta & 0 & 0 \\ 
-\Delta & -\kappa & \gomC & 0 \\ 
\hline
\rule{0pt}{2.8ex} 0 & 0 & -\frac{\Gamma_{m}}{2} & -\omega_{m} \\
\gomC & 0 & \omega_{m} & -\frac{\Gamma_{m}}{2}
\end{array} \right)\,\,.
\end{eqnarray}
The frequency $\Delta=\omega_{p}-\omega_{c}^{\prime}$ corresponds to the pump detuning from cavity resonance ($\omega_{c}^{\prime}$ includes the static shift contained in $H_{\mathrm{static}}$), while $\gomC=2\gom\sqrt{n}$ defines the effective optomechanical coupling rate. Photonic and phononic perturbative inputs (outputs) are symbolized by $\hat{\alpha}_{\inin}$ ($\hat{\alpha}_{\out}$) and $\hat{\eta}_{\inin}$ ($\hat{\eta}_{\out}$), respectively. Operator $\hat{\alpha}_{\inin}$ ($\hat{\alpha}_{\out}$) groups inputs (outputs) from the left and right ends of the cavity, $\hat{\alpha}_{\inL}$ ($\hat{\alpha}_{\outL}$) and $\hat{\alpha}_{\inR}$ ($\hat{\alpha}_{\outR}$), respectively, as well as from loss channels $\hat{\alpha}_{\inV}$ ($\hat{\alpha}_{\outV}$): 
\begin{equation}
\sqrt{\gamma_{T}}\,\hat{\alpha}_{\inin}=\sqrt{\gamma_{L}}\,\hat{\alpha}_{\inL} + \sqrt{\gamma_{R}}\,\hat{\alpha}_{\inR} + \sqrt{\gamma_{V}}\,\hat{\alpha}_{\inV}\,\,.
\end{equation}
The optical damping rates through each port, $\gamma_{L}$, $\gamma_{R}$ and $\gamma_{V}$, collectively define the cavity half-linewidth $\kappa$ as  $2\kappa=\gamma_{L}+\gamma_{R}+\gamma_{V}$. Analogously, communication between the mechanical oscillator and its environment takes place at a rate $\Gamma_{m}$.

Eqs.~(\ref{EomTime}--\ref{M}) 
show that the oscillator momentum is susceptible to fluctuations in the intracavity photon number, while the induced phase shift on circulating photons is dependent on the oscillator position. The mutual optomechanical transduction, captured by the off-diagonal block matrix $\mathbf{T}$, leads to closed-loop feedback. The standard formalisms from control theory thus provide appropriate means to construct a complete yet simple model for optomechanical systems. 

First, Eqs.~(\ref{EomTime}--\ref{M}) 
can be translated into a time-domain block diagram, as shown in Fig.~\ref{fig:BlockDiagram}(a). 
Such representations are analogous to earlier flow-chart depictions of optomechanical interactions \cite{Tsang2010}. Optomechanical interactions are shown as vertical lines of communication in Fig.~\ref{fig:BlockDiagram}, 
between the otherwise freely evolving optical and mechanical fields. 

Second, insight into the gains of the optomechanical feedback loop can be obtained by studying the evolution of field operators in frequency space, translating Eq.~(\ref{EomTime}) into  a pair of governing equations\footnote{The following Fourier transform convention is used: $\tilde{f}(\omega)=\int_{-\infty}^{\infty}f(t)e^{i\omega\,t}dt$}:
\begin{eqnarray}\label{EomFreq1}
\left( \begin{array}{c}
\hatt{a}_{+} \\
\hatt{a}_{-}
\end{array} \right)&=&\Fa\left\lbrack
\mathbf{T}
\left( \begin{array}{c}
\hatt{z} \\
\hatt{p}
\end{array} \right) + 
\sqrt{\gamma_{T}}
\left( \begin{array}{c}
\hatt{\alpha}_{\inin+} \\
\hatt{\alpha}_{\inin-}
\end{array} \right)\right\rbrack\,\,,\\ \nonumber
\left( \begin{array}{c}
\hatt{z} \\
\hatt{p}
\end{array} \right)&=&\Fb\left\lbrack
\mathbf{T}
\left( \begin{array}{c}
\hatt{a}_{+} \\
\hatt{a}_{-}
\end{array} \right) + 
\sqrt{\Gamma_{m}}
\left( \begin{array}{c}
\hatt{\eta}_{\inin+} \\ 
\hatt{\eta}_{\inin-} 
\end{array} \right)\right\rbrack\,\,,
\end{eqnarray}
where
\begin{eqnarray}\label{TransferFunctions}
\Fa &=&
\frac{1}{(\kappa-i\omega)^{2}+\Delta^{2}}
\left( \begin{array}{cc}
\kappa-i\omega & \Delta \\ -\Delta & \kappa-i\omega
\end{array} \right) \,\,, \\
\Fb &=&
\frac{1}{(\frac{\Gamma_{m}}{2}-i\omega)^{2}+\omega_{m}^{2}}
\left(\begin{array}{cc}
\!\frac{\Gamma_{m}}{2}-i\omega\! & \!-\omega_{m}\! \\ 
\nonumber
\!\omega_{m}\! & \!\frac{\Gamma_{m}}{2}-i\omega\! \end{array}\right)\!.
\end{eqnarray}

In Eqs.~(\ref{EomFreq1}), the intracavity photon (phonon) field is the sum of an optomechanical transduction of the intracavity phonon (photon) field and a cavity- (oscillator-) induced filtering of the $+$ and $-$ optical (mechanical) input quadratures, captured by $\Fa$ ($\Fb$). Eqs.~(\ref{EomFreq1}) are collectively represented by the block diagram shown in Fig.~\ref{fig:BlockDiagram}(b). 

By virtue of our approximations, the optomechanical amplifier model considered here is entirely linear; amplifier inputs and outputs can be completely parametrized by a set of transfer matrices. In typical optomechanics systems, only the optical field is detectible, so we will only give those transfer matrices that connect the inputs to the optical output.  A key result of this paper is that many of the salient features of optomechanics depend only on these transfer matrices, and not on the specific nature of the drive fields or, among other possibilities, 
on the assorted experimental configurations by which many phenomena of optomechanical systems have been uncovered.  The optical output is
\begin{equation}\label{EomIntraCav}
\ftmatrix{a} = \Haalph \sqrt{\gamma_T} \ftmatrixIn{\alpha} + \Haeta \sqrt{\Gm} \ftmatrixIn{\eta}. 
\end{equation}
The transfer matrices $\Haalph$ and $\Haeta$ have units of inverse frequency, and can be calculated by solving Eq.~(\ref{EomFreq1}).  The result is
\begin{align}
\label{eqn:Haalph}
\Haalph &= \begin{pmatrix} 1+\Hpalph & 0 \\ \Hmalph & 1 \end{pmatrix}  \Fa, \\ 
\label{eqn:Haeta}
\Haeta &= \begin{pmatrix} \Hpalph & 0 \\ \Hmalph & 0 \end{pmatrix} \begin{pmatrix} \Hpeta & \Hmeta \\ 0 & 0 \end{pmatrix}.
\end{align}
The individual matrix elements can be related to a common optomechanical gain $G$.  For systems with high mechanical quality factors, where the damping-induced frequency pulling of the oscillator may be neglected, one finds:
\begin{align}
\label{eqn:G}
G &= \frac{-s(\w)}{\wm^2 + s(\w) - \w^2 -i\w (\Gm \! + \! \Gopt(\w))},
\end{align}
\begin{align}
\label{eqn:Hpmalph}
\Hpalph &= G, & \Hmalph &= G\,\frac{\k-i\w}{\D}\,\,, \\
\label{eqn:Hpmeta}
\Hpeta &= -\frac{1}{g_C}\frac{\Gm/2-i\w}{\wm}\,\,, &
\Hmeta &= \frac{1}{g_C}\,\,.
\end{align}

In the above, $s(\w)$ represents the stiffening of the oscillator due to optomechanics, and $\Gopt(\w)$ represents an optomechanically induced damping rate:
\begin{align}
s(\w) &= \frac{\wm \D g_C^2}{\k^2 + \D^2 - \w^2}, \\
\Gopt(\w) &= \frac{2\k}{\k^2 + \D^2 - \w^2}(\wm^2  -\w^2).
\end{align}
The stiffening shifts the oscillator's resonant frequency to a value
\begin{align}
\nonumber \we^2 &= \frac 12 (\k^2 \mathord +\D^2 \mathord +\wm^2) - \frac 12 \sqrt{(\k^2 \mathord  + \D^2  \mathord  -\wm^2)^2  \mathord - 4 \wm \D g_C^2} \\
&\approx \wm^2 + s(\wm),
\end{align}
where the approximation holds for small shifts of $\we$ from $\wm$.

The properties of the amplifier can be parameterized by the optomechanical cooperativity, defined as $\Copt=\gomC^2/(\kappa\Gamma_{m})$. The cooperativity allows comparison between optomechanical systems over a broad range of parameters. However, $\Copt$ does not contain information about the detuning from cavity resonance. To draw comparison between systems operating away from cavity resonance, it is more convenient to adopt an ``optomechanical damping parameter'', defined as $\Dopt=\Gamma_{\opt}(\we)/\Gamma_{m}$.


In line with our earlier assumptions, Eq.~(\ref{EomIntraCav}) holds only for linear and stable systems. In optomechanics, this condition is always satisfied when the 
optomechanically induced cavity resonance shift is small compared to the cavity linewidth ($g_C < \kappa$) and when pumping to the red of cavity resonance ($\Delta \leq 0$), where backaction cooling dominates \cite{Cohadon1999, Mancini1998, Marquardt2006}; parametric instability can occur when pumping to the blue of cavity resonance \cite{Braginsky2002b, Marquardt2006, Kippenberg2005, Arcizet2006}. Eq.~(\ref{EomIntraCav}) is consequently invalid under optical bistability, where the pump detuning can spontaneously transition between negative and positive values due to changes in the instantaneous cavity resonance frequency.

Eq.~(\ref{EomIntraCav}) represents cavity optomechanics as a phase-sensitive amplifier \cite{Caves1982, Haus1962}. The amplifier has intrinsic quantum noise due to zero-point optical and mechanical fluctuations entering into the cavity. Furthermore, the amplifier can exhibit gain below unity in certain quadratures and frequency bands. This feature leads to a reduction of optical shot-noise below the level of Poissonian statistics \cite{Fabre1994, Borkje2010, Heidmann1997}, as shown in following sections.

\section{Intracavity Response}\label{sect:Intra}

The amplifier model developed above is applied here to study the response of the intracavity field to both optical and mechanical drives. Since the amplifier is linear, its susceptibility to each input type is completely decoupled, and we therefore treat the response to each input separately. 
In \S\ref{sect:IntraOpt}, we consider the transduction of the optical field, and show that OMIT and ponderomotive squeezing both arise from the same physics.  In \S\ref{sect:IntraMech}, we examine the transduction of mechanical inputs onto the optical field.

\subsection{Response to Optical Inputs - Ponderomotive Attenuation and OMIT}\label{sect:IntraOpt}

The amplifier's response to optical inputs is characterized by the optical transfer matrix $\mathbf{H}_{\alpha}$, given in Eqs.~(\ref{eqn:Haalph}--\ref{eqn:Hpmalph}).  This transfer matrix can be used to predict experimental outcomes. Here, we consider two types of experiments.  In the first, modulation is applied to the pump (this applied modulation may be a coherent drive, classical noise, or quantum fluctuations), and the response of the optomechanical system is detected in the resulting amplitude or phase modulation of the optical field.  Such experiments have demonstrated both ponderomotive squashing \cite{Marino2010, Verlot2010} (attenuation of classical noise) and squeezing \cite{Brooks2011} (attenuation of quantum noise), as well as the transduction of coherent signals \cite{Verlot2010, Brooks2011}.  In the second type of experiment, 
a weak probe at a single frequency accompanies the pump into the cavity and is detected, yielding observations of OMIT and optomechanically induced amplification \cite{Safavi2011, Weis2010, Huang2011, Huang2011b}.  We show that both types of experiments are related via a simple unitary rotation of the transfer matrix.

\begin{figure}
	\centering
		\includegraphics{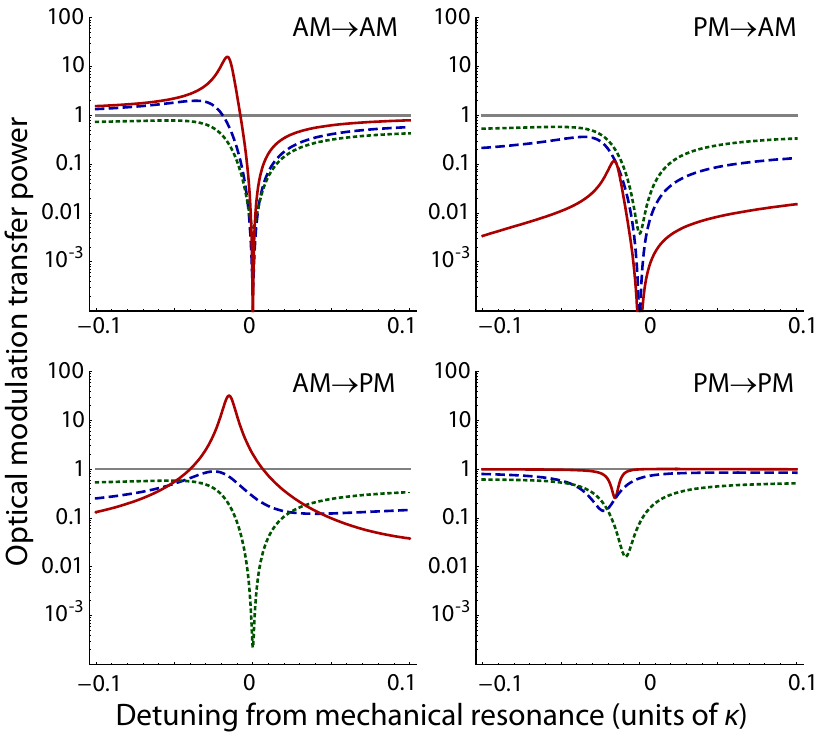}
	\caption{(Color online) Square magnitude of the elements of the optical modulation transfer matrix. The elements describe the transduction of amplitude and phase modulation from input to output. Powers are normalized by the total intracavity attenuation in the absence of optomechanics ($G=0$). Each element is plotted for the unresolved ($\wm/\k=0.2$, solid red), intermediate ($\wm/\k=1$, dashed blue), and resolved ($\wm/\k=5$, dotted green) sideband cases, with the anti-Stokes mechanical sideband ($\D+\wm$) fixed at $-0.5~\k$ from cavity resonance. 
Ponderomotive attenuation is observed in the vicinity of $\w=\wm$, while ponderomotive amplification is prevalent only for the unresolved sideband case, in the vicinity of $\w=\we$. For all plots, $\Dopt = 30$ and $Q = 1000$.  
	}
	\label{fig:HaMod}
\end{figure}

The first type of experiment is characterized by the transduction of AM and PM from an optical input to the intracavity optical field.  Because the static field $a_0$ is real, amplitude and phase modulation quadratures correspond identically to the $\hat{a}_\pm$ observables.  However, because of the cavity phase rotation, the input field $\alpha_{\mathrm{in},0}$ has a complex phase $\psi_c = \arctan(\D/\k)$
, and the observables corresponding to AM and PM are
\begin{align}
\ftmatrixAP{\alpha} &= \mathbf{R}(-\psi_c) \ftmatrixIn{\alpha}\,\,,
\end{align}
where $\mathbf{R}$ is a rotation matrix defined as
\begin{align}
\mathbf{R}(\theta) = \begin{pmatrix} \cos\theta & -\sin\theta \\ \sin\theta & \cos\theta \end{pmatrix}\,\,.
\end{align}
The input-output relation is thus given by the modulation transfer matrix $\HaMT$ as
\begin{align}
\ftmatrix{a} &= \Haalph  \mathbf{R}(\psi_c)\sqrt{\gamma_T} \ftmatrixAP{\alpha}\,\,, \\
\nonumber
&= \HaMT \sqrt{\gamma_T} \ftmatrixAP{\alpha}\,\,.
\end{align}

Fig.~\ref{fig:HaMod} shows the power (square magnitude) of each element of $\HaMT$.  For a pump beam detuned to the red of cavity resonance, one sees an attenuation of modulation in the vicinity of $\w=\wm$ in the output AM spectra
, caused by the destructive interference of the input fluctuations with the mechanically transduced fluctuations.  When the system is driven by optical shot noise, this attenuation gives ponderomotive squeezing of the optical field (see \S\ref{sec:drive:vac}).  In the unresolved sideband case ($\wm\ll\k$), one also observes amplification of transduced AM 
near $\w=\we<\wm$ due to constructive feedback.  As $\wm$ is increased above $\k$, the optomechanical damping in the vicinity of $\we$ increases, broadening and attenuating the amplified peak.

\begin{figure} 
\centering
	\includegraphics [width= 2.75in] {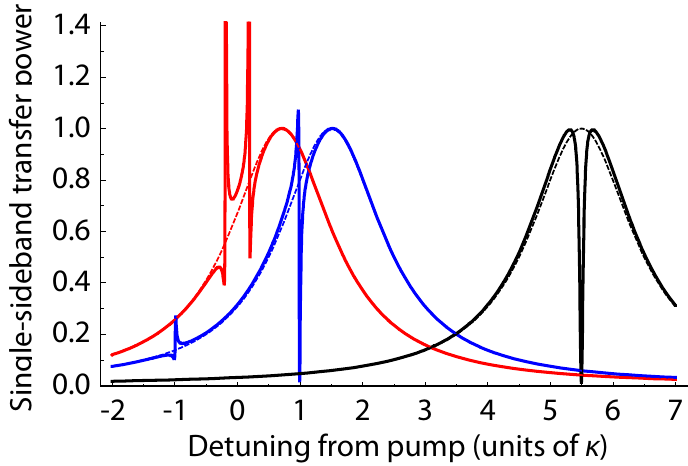} 
	\caption{(Color online) Sum of the square magnitude of the $\hatt{\alpha}(\w)\rightarrow \hatt{a}(\w)$ term in the single-sideband spectral transfer matrix $\HaSS$. Probe frequencies $\omega$ are quoted in units of $\kappa$ relative to the pump frequency at $\omega=0$. Three cases are considered: unresolved ($\wm/\k=0.2$, red), intermediate ($\wm/\k=1$, 
blue) and resolved ($\wm/\k=5$, 
black) sideband regimes, with the anti-Stokes mechanical sideband ($\D+\wm$) fixed at $-0.5~\k$ from cavity resonance for the first two cases, and fixed on cavity resonance for the resolved sideband case. 
Solid lines indicate responses in the presence of an oscillator, while dashed lines apply to empty cavities. No normalization is performed (1 represents unity gain). For all plots, $\Dopt = 30$ and $Q = 1000$. 
}
	\label{fig:OMITSpectra}
\end{figure}

For the second experiment, we calculate the transduction of a single-tone probe $\hatt{\alpha}(\w)$.  
The single-sideband transfer matrix relates the input and output of pure tones at $\pm \w$:
\begin{align}
\begin{pmatrix} \hatt{a}(\w) \\ \hatt{a}^\dagger(-\w) \end{pmatrix} = \HaSS \sqrt{\gamma_T} \begin{pmatrix} \hatt{\alpha}(\w)\\ \hatt{\alpha}^\dagger(-\w) \end{pmatrix}\,\,,
\end{align}
\vspace{-\baselineskip}
\begin{align}
\nonumber
\HaSS &= \mathbf{U} \Haalph \mathbf{U}^{-1}\,\,, & 
\mathbf{U} &= \frac{1}{\sqrt{2}} \begin{pmatrix} 1 & -i \\ 1 & i \end{pmatrix}\,\,.
\end{align}

\noindent Fig.~\ref{fig:OMITSpectra} shows the sum of the square magnitude of the $\HaSS$ terms that link the pure tone input $\hatt{\alpha}(\w)$ to the corresponding intracavity field $\hatt{a}(\w)$. It reflects the total intensity transduced in this single optical sideband (\textit{i.e.}~quadrature-insensitive detection). In the resolved-sideband limit, with the pump anti-Stokes mechanical sideband centered on the cavity resonance ($\omega_{m}=-\Delta$), a prominent dip at $\omega_{m}$ is visible. This feature is the hallmark of OMIT \cite{Safavi2011, Weis2010}.

This resolved-sideband probe spectrum changes when transitioning to the unresolved-sideband limit and detuning the pump anti-Stokes sideband from cavity resonance ($\omega_{m}<-\Delta$), with increased amplification at the effective oscillation frequency $\we$ and decreased attenuation at $\wm$.

Our linear feedback model reveals clearly that ponderomotive attenuation and OMIT originate from the same physical phenomenon. Moreover, the model highlights that OMIT can be reinterpreted as a classical analog of ponderomotive squeezing, in that it demonstrates that certain matrix elements of the optical-to-optical transfer matrix have magnitude well below unity.

\subsection{Response to Mechanical Inputs}\label{sect:IntraMech}

\begin{figure}
	\centering
		\includegraphics{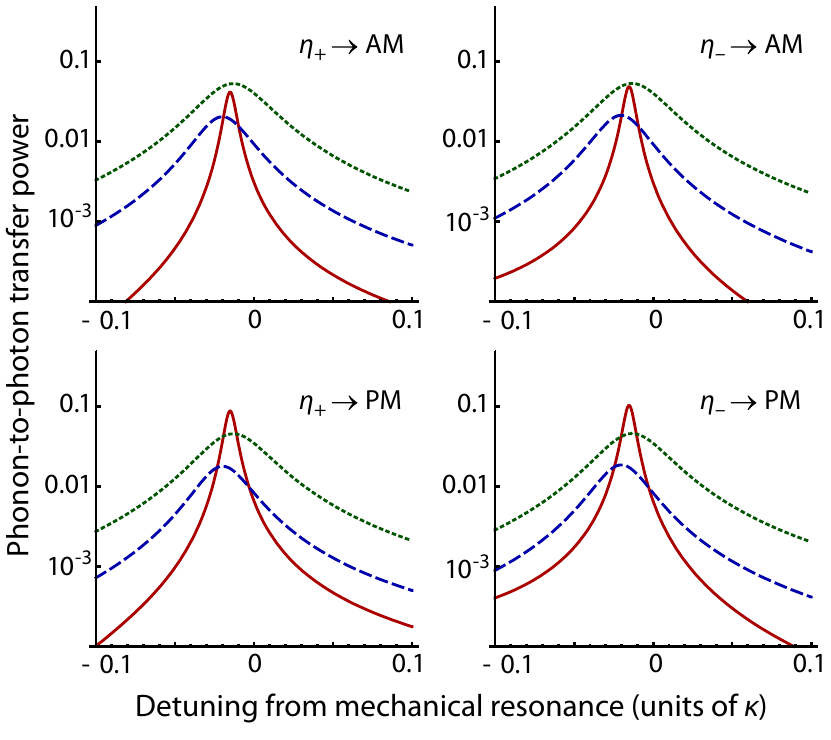}
	\caption{(Color online) Square magnitude of elements of the mechanical $\rightarrow$ optical modulation transduction matrix $\sqrt{\Gm} \Haeta$.  The ordinate axis is in units of square modulation quanta of the intracavity field $\hat{a}$ per square flux quanta of the mechanical input field $\hat{\eta}$.  Each element is plotted for the unresolved ($\wm/\k=0.2$, solid red), intermediate ($\wm/\k=1$, dashed blue), and resolved ($\wm/\k=5$, dotted green) sideband cases, with the anti-Stokes mechanical sideband ($\D+\wm$) fixed at $-0.5~\k$ from cavity resonance. 
For all plots, $\Dopt = 30$ and $Q = 1000$.}
	\label{fig:HaMechMax}
\end{figure}

Next, the transduction of mechanical inputs onto the circulating light field is studied.  Fig.~\ref{fig:HaMechMax} shows the power in each element of the mechanical to optical transduction matrix $\Haeta$, multiplied by the input phonon coupling $\Gm$.  Expressed in this manner, the matrix elements give the optomechanical conversion of flux of phonon quanta into quanta of optical modulation.  Transduction of $\eta_-$ into PM follows a Breit-Wigner function, with response peaked at the effective oscillation frequency $\we$.  The AM response to $\eta_{-}$ is scaled by $\D/(\k-i\w)$ relative to that of PM, and the transduction of $\eta_+$ into AM/PM carries an additional factor of $(i\w-\Gm/2)/\wm$ relative to $\eta_-$.


\section{Post-Cavity Detection}\label{sect:BalancedDetection}
We now extend the model developed in section \ref{sect:Model} to consider the cavity output field. Photons 
exiting the cavity through the right and left mirrors 
form the new outputs of the optomechanical amplifier. 

Transfer functions connecting optical and mechanical inputs to the optomechanically colored cavity output light follow from Eq.~(\ref{EomIntraCav}) according to cavity boundary conditions \cite{Walls1995}.  For any of the ports $l=L,\:R$, or $V$, the boundary condition is
\begin{equation}
\hat{\alpha}_{\mathrm{in},l} + \hat{\alpha}_{\mathrm{out},l} = \sqrt{\gamma_l} \, \hat{a}.
\end{equation}
The boundary condition leads to expressions for the transfer matrix connecting any input $\hat{\alpha}_{\inin,k}$ (where $k$ = $L$, $R$, $V$, or $\eta$) to light in output field $\hat{\alpha}_{\out,j}$ (where $j$ = $L$ or $R$):
\begin{align}
\label{EqnExtraCav}
\mathbf H_{j\,k} &{\mathop{=}_{j\neq k}} \sqrt{\gamma_j \gamma_k}\, \mathbf H_{k}, & \mathbf H_{j\,j} &= \gamma_j \Haalph - \mathbf{1},
\end{align}
where $\mathbf H_{k} = \Haalph\textrm{ or }\Haeta$ for $k=L,R,V\textrm{ or }\eta$, respectively, and $\mathbf{1}$ is the identity matrix.

An optical signal measured in reflection thus includes beats between the exiting intracavity field and the reflected input beam.  In contrast, both the optical signals from transmitted light and from the mechanical input are determined by scaling the intracavity field (see \S\ref{sect:Intra}).  Both amplifier output ports can consequently carry distinct signatures of a common optomechanical interaction.

\begin{figure}
	\centering
		\includegraphics{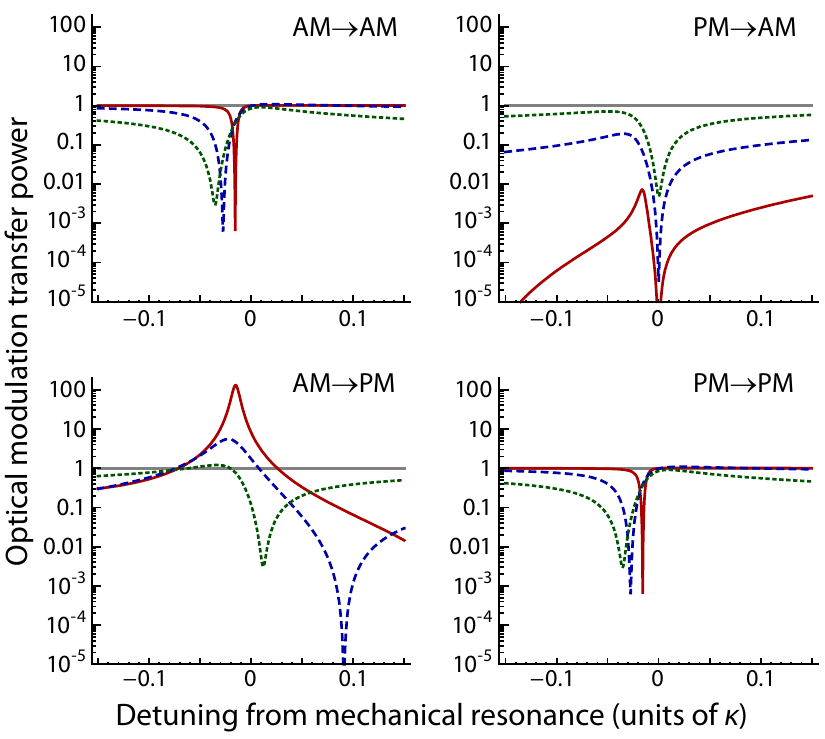}
	\caption{(Color online) Square magnitude of the elements of $\mathbf H^{\mathrm{MT}}_{j\,j}$, the optical modulation matrix connecting inputs and outputs in reflection. The elements describe the transduction of amplitude and phase modulation. Each element is plotted for the unresolved ($\wm/\k=0.2$, solid red), intermediate ($\wm/\k=1$, dashed blue), and resolved ($\wm/\k=5$, dotted green) sideband cases, with the anti-Stokes mechanical sideband ($\D+\wm$) fixed at $-0.5~\k$ from cavity resonance. 
	For all plots, $\Dopt = 30$ and $Q = 1000$.  
	}
	\label{fig:HaRefl}
\end{figure}

Let us apply these results to determine the output field response in reflection to applied pump modulations. As noted in \S\ref{sect:IntraOpt}, the entering and exiting fields carry separate complex phases relative to the static intracavity field $a_{0}$ due to cavity rotation. Both phase angles are linked by the cavity boundary condition. For a one-sided, lossless cavity, the phase angles differ by $\phi_c=-\arctan\left(2\k\D/\left(\k^2-\D^2\right)\right)$. 
Input AM and PM at port $j$ are transduced into observable AM and PM at the output of that same port according to
\begin{align}
\ftmatrixAPout{\alpha} &= \mathbf H^{\mathrm{MT}}_{j\,j} \ftmatrixAPin{\alpha}\,\,, \\
\nonumber
\mathbf H^{\mathrm{MT}}_{j\,j} &= \mathbf{R}(-\psi_c)\mathbf{R}(-\phi_c)\mathbf H_{j\,j}\mathbf{R}(\psi_{c})\,\,.
\end{align}
Fig.~\ref{fig:HaRefl} shows the power (magnitude square) of each element of $\mathbf H^{\mathrm{MT}}_{j\,j}$ for a one-sided, lossless cavity. The results are considerably different from those measured in transmission or intracavity (see Fig.~\ref{fig:HaMod}). The differences are due to the combined effect of cavity-induced phase rotation of the exiting intracavity field and its interference with light reflecting off the input/output mirror. 

\section{Specific input conditions}
\label{sec:drive}

We now turn to the optomechanical amplifier response to specific inputs: quantum fluctuations of the mechanical and optical fields, a thermal mechanical bath, and external forces.

To characterize the response, we determine the power spectral density (PSD) of an output field quadrature $\hat{X}_{\theta j}$.  The quadrature is defined by:
\begin{equation}\label{eq:quad}
\hat{X}_{\theta j} \equiv \hat{\alpha}_{\out,j+} \cos\theta + \hat{\alpha}_{\out,j-} \sin\theta\,\,.
\end{equation}
The PSD is defined as
\begin{align}
S_{\theta j}(\w) &\equiv \left \langle \left |\hatt{X}_{\theta j}\right |^2 \right \rangle \\
\nonumber
 &= \left \langle \left | \sum_k \begin{pmatrix} \cos\theta & \sin\theta \end{pmatrix} \mathbf{H}_{j\,k} \begin{pmatrix} \hatt{\zeta}_{\inin,+} \\ \hatt{\zeta}_{\inin,-} \end{pmatrix} \right |^2 \right \rangle\,\,,
\end{align}
where $\zeta=\alpha\textrm{ or }\eta$ for $k=L,R,V\textrm{ or }\eta$, respectively. If the inputs are incoherent, such that $\langle \hatt{\zeta}_{\inin1}^{\dagger}\hatt{\zeta}_{\inin2} \rangle = 0$ along two distinct input ports $k_1$ and $k_2$, the PSD can be rewritten as a sum of transduced powers:
\begin{align}
S_{\theta j}(\w) = & \sum_k \left \langle \left | \begin{pmatrix} \cos\theta & \sin\theta \end{pmatrix} \mathbf{H}_{j\,k} \begin{pmatrix} \hatt{\zeta}_{\inin,+} \\ \hatt{\zeta}_{\inin,-} \end{pmatrix} \right |^2 \right \rangle \\ &\textrm{(assuming incoherent inputs)}\,\,.\nonumber 
\end{align}

Finally, the PSD of a real observable (\textit{e.g.}~$X_{\theta}$) is necessarily symmetric in frequency, so we quote the  symmetrized PSD
\begin{equation}
S_{\theta j,S}(\w) \equiv \frac{1}{2} \left ( S_{\theta j} (\w) + S_{\theta j} (-\w) \right )\,\,.
\label{eq:sym}
\end{equation}

\begin{figure*}[!t]
	\centering
		\includegraphics{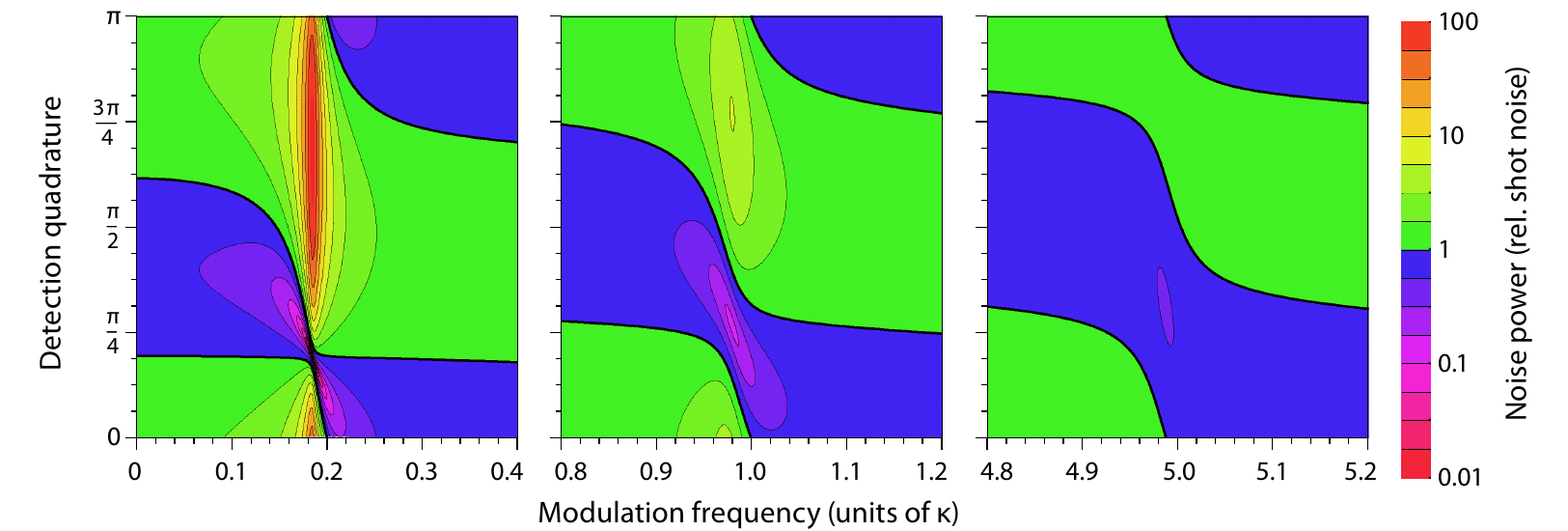}
	\caption{(Color online) Ponderomotive amplification and squeezing.  Each panel shows the noise power (relative to shot-noise) of the optical field vs. frequency and quadrature for a one-sided lossless cavity.  Quadrature angles refer to $\theta$ in Eq.~(\ref{eq:quad}) ($\theta=0,\pi/2$ correspond to $+\textrm{,}-$ quadrature, respectively). Columns show, respectively, the unresolved ($\wm/\k = 0.2$), intermediate ($\wm/\k = 1$), and resolved ($\wm/\k = 5$) sideband cases, with the anti-Stokes mechanical sideband ($\D+\wm$) fixed at $-0.5~\k$ from cavity resonance. For all plots, $\Dopt=30$ and $Q=1000$.}
	\label{fig:squeeze}
\end{figure*}

\subsection{Optical and Mechanical Vacuum Fluctuations}
\label{sec:drive:vac}

Applying quantum perturbations to drive the oscillator allows the quantum nature of optomechanics to be probed. We study the case of quantum drives here by considering a shot-noise-limited pump input, where the cavity noise spectrum is dominated by optical and mechanical vacuum fluctuations. Under such a premise, optomechanical interactions imprint these zero-point fluctuations on the circulating light field with gain. Regions where the resulting fluctuations drop below the standard quantum limit (SQL) yield quadrature-squeezed light, \textit{i.e.}~ponderomotive squeezing \cite{Fabre1994}. Such squeezing is particularly important for gravitational-wave detectors, such as LIGO, where squeezed light sources provide enhanced detection sensitivity 
\cite{Caves1981, Harms2003}. Ponderomotive squeezing serves as an in-situ means 
to produce squeezed light at low frequencies, where detectors are sensitive to gravitational-wave perturbations \cite{Corbitt2006b}. A first experimental observation of ponderomotive squeezing has recently been reported \cite{Brooks2011}.

We start with the noise spectrum of light reflected from a one-sided lossless cavity ($\gamma_R\mathbin=2\k$, $\gamma_L\mathbin=\gamma_V\mathbin=0$). The general expression for the symmetrized PSD relative to shot-noise in this ideal case is
\begin{equation}
\bar{S}^{\ideal}_{\theta R,S}(\w)=\sum_{k=R,\eta} \left \| \begin{pmatrix} \cos\theta & \sin\theta \end{pmatrix} \mathbf{H}_{R\,k} \right \|^2\,\,.
\end{equation}
The overbar on $\bar{S}^{\ideal}_{\theta R,S}$ specifies that the symmetrized PSD is normalized by shot-noise and is therefore dimensionless. In addition, $\|\dots\|^2$ represents the vector inner product.

$\bar{S}^{\ideal}_{\theta R,S}$ is shown in Fig.~\ref{fig:squeeze} for the case of sideband cooling ($\D<0$). With the anti-Stokes mechanical sideband ($\D+\wm$) held fixed at $-0.5~\k$, ponderomotive squeezing is found to be largest near the $\theta=\pi/4$ quadrature at frequencies around $\we$. For each sideband regime, the maximum squeezing quadrature and frequency are set by the competition of optomechanically squeezed vacuum inside the cavity and uncorrelated vacuum reflecting off the transmissive mirror. The suppression of quantum noise below the SQL results from the oscillator responding out of phase to vacuum perturbations above its effective mechanical resonance $\we$. Quantum optical fluctuations are also mechanically transduced between uncorrelated conjugate quadratures (\textit{e.g.}~AM $\leftrightarrow$ PM). This leads to shot-noise amplification at all frequencies in the conjugate quadrature to that of maximum squeezing (near $\theta=3\pi/4$ in Fig.~\ref{fig:squeeze}).

For a two-sided or lossy cavity, shot noise reflecting off the output port adds incoherently with vacuum fluctuations entering the cavity from other input ports. In addition, loss channels replace colored vacuum exiting the cavity with uncorrelated vacuum. Defining the relative photon extraction efficiency through the output port as $\varepsilon_{\out}$ ($\gamma_R\mathbin=2\k\,\varepsilon_{\out}$, $\gamma_L+\gamma_V\mathbin=2\k\,(1-\varepsilon_{\out})$), 
and expressing the fraction of cavity output field detected as $\varepsilon_{\detec}$, one finds that the detected spectrum of noise relative to shot-noise is \cite{Walls1995}
\begin{equation}\label{Seff}
\bar{S}^{\mathrm{obs}}_{\theta R,S}(\omega)=\varepsilon_{\tot} \bar{S}^{\mathrm{ideal}}_{\theta R,S}(\omega)+1-\varepsilon_{\tot}\,\,,
\end{equation}
where $\varepsilon_{\tot}=\varepsilon_{\out}\cdot\varepsilon_{\detec}$. If measurements are to be made on only one output port, any departure from the ideal one-sided case therefore results in a reduction of observable squeezing.

\begin{figure*}[!t]
	\centering
		\includegraphics{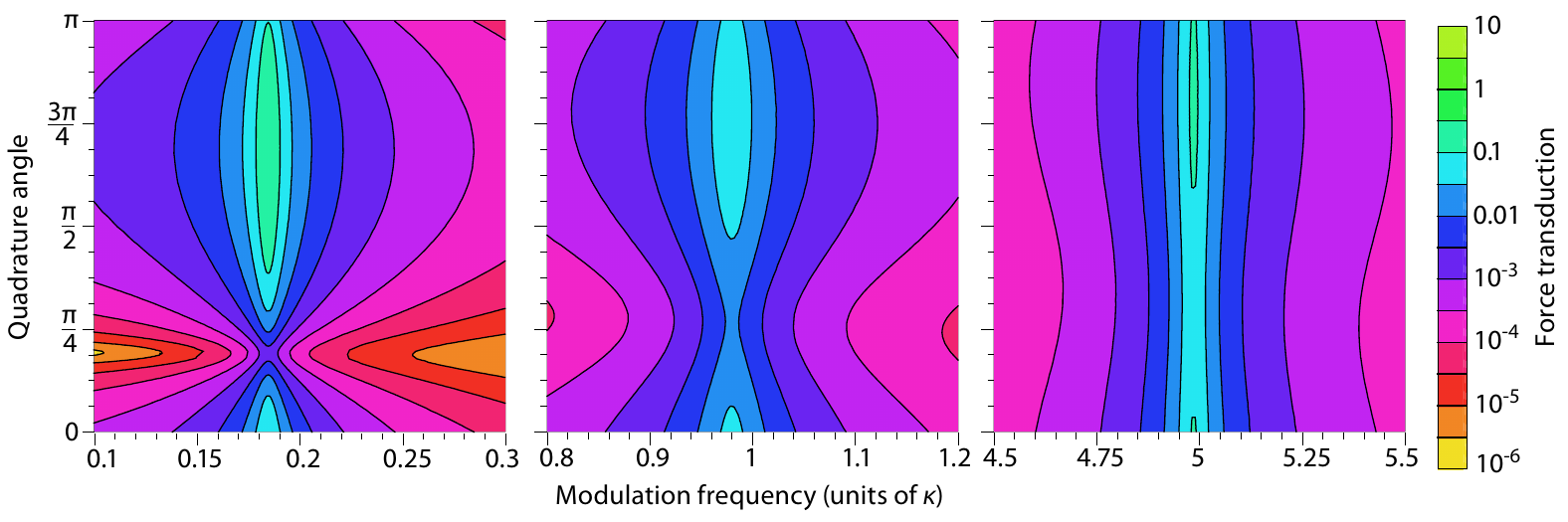}
	\caption{(Color online) External force sensitivity. Figure shows the optical power spectrum from external force transduction normalized by the drive power vs. frequency and quadrature. Quadrature angles refer to $\theta$ in Eq.~(\ref{eq:quad}) ($\theta=0,\pi/2$ correspond to $+\textrm{,}-$ quadrature, respectively). Columns show, respectively, the unresolved ($\wm/\k = 0.2$), intermediate ($\wm/\k = 1$), and resolved ($\wm/\k = 5$) sideband cases, with the anti-Stokes mechanical sideband ($\D+\wm$) fixed at $-0.5~\k$ from cavity resonance.  For all plots, $\Dopt = 30$ and $Q = 1000$.}
	\label{fig:ForceTrans}
\end{figure*}

\subsection{Mechanical Drive}
\label{sec:drive:mech}

Several noise models for mechanical disturbances have been proposed \cite{WilsonRae2008, Hu1992, Caldeira1983}. As done in earlier works \cite{Fabre1994, Borkje2010}, here we study the Caldeira-Leggett model \cite{Caldeira1983} under the Markov approximation, where
\begin{equation}\label{ThermalModelTime}
\langle\hat{\eta}^{\dagger}_{\inin}(t)\hat{\eta}_{\inin}(t')\rangle=n_{\therm}\delta(t-t')\,\,,
\end{equation}
and $\Gamma_{m}$ is constant across frequencies. The term $n_{\therm}=[e^{\hbar\omega_{m}/(k_{\mathrm{B}}T)}-1]^{-1}$ corresponds to the mean phonon number, determined by the mechanical bath temperature $T$. This model is appropriate for oscillators with high mechanical quality factor $Q=\omega_{m}/\Gamma_{m}$. Under this model, the total output noise due to mechanical fluctuations relative to mechanical shot-noise is
\begin{equation}
S^\mathrm{mech}_{\pm R,S} = \frac{2\varepsilon_{\tot}}{\Copt f_\mathrm{BW}} \frac{\wm^2+\w^2}{\wm^2} (2 n_\therm + 1)|H_{\pm \alpha}|^2,
\end{equation}
where $f_\mathrm{BW}$ is the Fourier-transform bandwidth.

The drive source may also be an external force acting on the oscillator (\textit{e.g.}~a weak force that one wishes to detect via optomechanics). Since forces impart a change in momentum, a coherent force acting on the oscillator $\tilde{F}_{\ext}=F_{\ext}/f_\mathrm{BW}$ is only related to the quadrature input term $\hat{\tilde{\eta}}_{\inin-}$ as
\begin{equation}
\sqrt{\Gamma_{m}}\,\left\langle\hat{\tilde{\eta}}_{\inin-}\right\rangle =\frac{\tilde{F}_{\ext}}{p_{\ho}}\,\,.
\end{equation}
The term $p_{\ho}=\sqrt{\hbar\,M\omega_{m}/2}$ symbolizes the harmonic oscillator momentum of the mechanical resonator, where $M$ is the resonator's mass. 

The push provided by $\tilde{F}_{\ext}$ is imprinted on the optical pump field via the mechanically-applied phase shift (see Fig.~\ref{fig:BlockDiagram}): optical sidebands at the drive frequencies are promoted.  The optical power spectrum on the cavity output field arising from an external force, in units of square optical quanta per square force quanta, is given by 
\begin{equation}
\frac{S^\mathrm{ext}_{\theta R,S}}{|F_\mathrm{ext}|^2} = \frac{2\varepsilon_{\tot}}{\Copt} \left | \cos\theta + \frac{\k-i\w}{\D} \sin\theta \right |^2 \frac{|G|^2}{\Gm f_{\mathrm{BW}}^{2} p_\mathrm{HO}^2}
\end{equation}
Fig.~\ref{fig:ForceTrans} shows that the force transduction is peaked at the same frequency ($\w=\we$) and quadrature (near $\theta=3\pi/4$ when $\D+\wm$ is fixed at $-0.5~\k$) as that of maximum shot-noise amplification (see \S\ref{sec:drive:vac}). The maximization of force detection signal and quantum noise at the same point in parameter space motivates the question of optimal force sensing quadrature and condition.

\begin{figure}[h]
	\centering
		\includegraphics{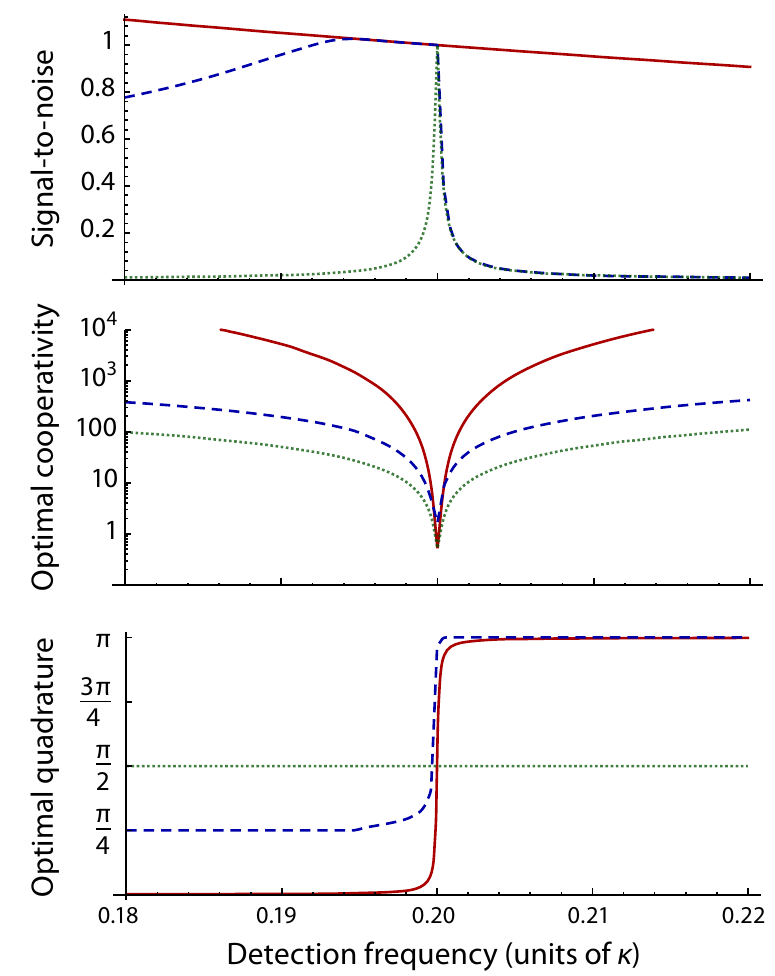}
	\caption{(Color online) Optimal force sensing with and without ponderomotive squeezing.  The top panel shows the prefactor for the optimal signal-to-noise (relative to that of Eq.~\ref{RextMax}) vs.\ force frequency, the middle panel shows $\Copt$ necessary to obtain optimal force sensing, and the bottom panel shows the necessary quadrature angle.  Each plot shows the case without squeezing ($\D=0$, solid red line) and with squeezing ($\D=-\k$, dashed blue line).  The optimal signal-to-noise ratio for the canonical force detection experiment at $\D=0$ and $\theta=\pi/2$ is included for reference (\textit{i.e.}~extension of  Eqs.~(\ref{RextMax}--\ref{Copt}) over all frequencies, dotted green line). 
	For all plots, $\wm/\k=0.2$ and $Q = 1000$.}
	\label{fig:RSNOpt}
\end{figure}

Temperature and force sensing have standard quantum limits (SQL) set by two competing optomechanical effects: the transduction of external forces onto the light field and the optomechanical gain of zero-point fluctuations.  SQLs for the canonical experiment of detection at $\D=0$, $\w=\wm$, $\theta=\pi/2$ and $\varepsilon_{\tot}=1$ have previously been derived \cite{Caves1982, Clerk2010}. The quantum limit on signal-to-noise ratio (SNR) for external force detection is reached when the cooperativity is tuned such that the noise arising from backaction (transduction of optical noise along the $\theta=0$ quadrature) is equal in magnitude to optomechanically colored optical shot noise along the $\theta=\pi/2$ quadrature.  Rederived using the linear amplifier model, the SNRs for sensing of thermal motion and sensing of coherent external forces under these conditions are given as 
\begin{align}\label{RextMax}
R_{\mathrm{SQL}}^\mathrm{therm} &= \left (1+\frac{3}{64}\frac 1{Q^2} +\mathcal{O}(Q^{-4}) \right ) n_\therm\,\,, \\
\nonumber R_{\mathrm{SQL}}^\mathrm{ext} &= \left (\frac 14 -\frac{5}{256}\frac{1}{Q^2} + \mathcal{O}(Q^{-4}) \right ) \frac{F^2_\mathrm{ext}}{\Gm f_\mathrm{BW} p_\mathrm{HO}^2},
\end{align}
obtained using an optomechanical cooperativity of
\begin{equation}\label{Copt}
C^\mathrm{SQL}_{\mathrm{opt}} = \frac12 \left ( 1+\frac{\wm^2}{\k^2} \right ) + \mathcal{O}(Q^{-2}).
\end{equation}
An adapted form of Eqs.~(\ref{RextMax}--\ref{Copt}) is shown in Fig.~\ref{fig:RSNOpt}.

We now extend this work to consider force detection at frequencies away from the mechanical resonance. The maximum SNR attainable in this case is given by
\begin{equation}
R_\mathrm{max}^{\mathrm{ext}} = \frac{1}{(\frac{1}{2Q})^2+\left(1+\left|\frac{\w}{\wm}\right|\right)^2}\:\frac{F^2_\mathrm{ext}}{\Gm f_\mathrm{BW} p_\mathrm{HO}^2}\,\,,
\end{equation}
using the quadrature angle and optomechanical cooperativity
\begin{align}
\theta^\mathrm{max} &= \arctan\left(\frac{\frac{1}{Q}\left|\frac{\w}{\wm}\right|}{(\frac{1}{2Q})^2+(1-\frac{\w^2}{\wm^2})}\right) \\[6pt]
\nonumber
\Copt^\mathrm{max} &= \frac{(1+\frac{\w^2}{\k^2})\left|\wm^2+\Gm^2/4-\w^2-i\,\Gm\w\right|^2}{2\,\Gm^2\wm\left|\w\right|},
\end{align}
These optimal solutions are shown in Fig.~\ref{fig:RSNOpt}. Results indicate that the optimal cooperativity quickly diverges to experimentally unrealistic values as the measurement frequency shifts away from mechanical resonance. In addition, we point out that these optimal solutions are extremely fragile to optical losses; $R_\mathrm{max}^{\mathrm{ext}}$ is significantly reduced at frequencies away from $\wm$ for $\varepsilon_{\tot} \lesssim 1$. 

Can we improve on the force transduction SNR by detuning off cavity resonance and benefitting from ponderomotive squeezing \cite{Corbitt2004, Kimble2001, Arcizet2006B}? We provide an answer to this question by applying a numerical optimization routine to identify $\Copt^\mathrm{max}$, $\theta^\mathrm{max}$ and $R_\mathrm{max}^{\mathrm{ext}}$ for $\wm/\k=0.2$ (unresolved sideband case) and various pump detunings $\D$, ranging from $-0.5~\k$ to $-6~\k$. 
The results under $\wm/\k=0.2$ and $\D/\k=-1$ are shown in Fig.~\ref{fig:RSNOpt}.  When comparing the signal to the total noise in the reflected optical field, we find that squeezing and detuning off cavity resonance \emph{do not} increase the signal-to-noise ratio above the maximum SNR found for $\D=0$. However, results also indicate that there is a range of frequencies ($\omega\lesssim\wm$) for which the optimal level of force sensitivity is attainable off cavity resonance with realistic cooperativities. \\



\section*{Conclusion}\label{Conclusion}

We have shown that linearized optomechanical systems can be modeled as linear optical amplifiers with mechanical and optical inputs. In this new context, ponderomotive squeezing, an entirely quantum effect, and OMIT are shown to be related: the latter is a classical manifestation of the former.  The optomechanical amplifier model was extended to predict observable power spectra in cavity output field under different drives. Our work indicates that ponderomotive squeezing is most visible in the unresolved-sideband limit. It also highlights that optical squeezing cannot be harnessed to surpass the maximum SNR for external force measurements on cavity resonance. 

The amplifier model offers a simple picture of optomechanical systems with insights into the sources of gain and lines of communication between the mechanical and optical degrees of freedom. As research in the field continues to diversify, we hope it may serve as a tool to model different setups under a common language and bridge separate concepts.

This work was supported by the AFOSR and NSF. T.B.\ acknowledges support from Le Fonds Qu\'eb\'ecois de la Recherche sur la Nature et les Technologies.

%

\end{document}